\documentclass[a4paper,fleqn,usenatbib]{mnras}

\usepackage[T1]{fontenc}
\usepackage{ae,aecompl}

\usepackage{ulem}
\usepackage{amssymb}

\usepackage{graphicx}
\usepackage{lscape}

\def\ergscm{erg~s$^{-1}$~cm$^{-2}$}

\def\lum{erg s$^{-1}$}

\def\apj{ApJ}

\def\aap{A\&A}

\def\apjs{ApJS}

\def\apjl{ApJLett}
\def\procspie{\ref@jnl{Proc.~SPIE}}   

\def\x1{NGC~5643~X-1}
\def\sn{X-1}
\def\nu{\textit{NuSTAR}}
\def\xmm{\textit{XMM--Newton}}

\title[\nu\ and \xmm\ observations of the ULX NGC 5643 X-1]{\nu\ and
  \xmm\ observations of the ultraluminous X-ray source NGC 5643 X-1} 
\author[Krivonos et al.]{Roman\,Krivonos$^{1}$\thanks{E-mail:
    krivonos@iki.rssi.ru} and Sergey\,Sazonov$^{1,2}$ 
\\
$^{1}$Space Research Institute of the Russian Academy of Sciences,
Profsoyuznaya Str. 84/32, 117997 Moscow, Russia\\
$^{2}$Moscow Institute of Physics and Technology, Institutsky per. 9,
141700 Dolgoprudny, Russia
}

\begin{document}
\label{firstpage}
\pagerange{\pageref{firstpage}--\pageref{lastpage}}
\maketitle

\begin{abstract}

We present a high-quality hard X-ray spectrum of the ultraluminous
X-ray source (ULX) NGC~5643 X-1 measured with \nu\ in May--June
2014. We have obtained this spectrum by carefully separating the
signals from the ULX and from the active nucleus of its host galaxy
NGC~5643 located 0.8 arcmin away. Together with long \xmm\ observations
performed in July 2009 and August 2014, the \nu\ data confidently
reveal a high-energy cutoff in the spectrum of \x1\ above
$\sim 10$~keV, which is a characteristic signature of ULXs. The \nu\ and
\xmm\ data are consistent with the source having a constant luminosity
$\sim 1.5\times10^{40}$~\lum\ (0.2--12~keV) in all but the latest
observation (August 2014) when it brightened to $\sim
3\times10^{40}$~\lum. This increase is associated with the dominant,
hard spectral component (presumably collimated emission from the inner
regions of a supercritical accretion disc), while an additional, soft
component (with a temperature $\sim0.3$~keV if described by multicolour
disc emission), possibly associated with a massive wind outflowing from the
disc, is also evident in the spectrum but does not exhibit significant
variability. 
 
\end{abstract}


\section{Introduction}

Ultraluminous X-ray sources (ULXs) are point-like X-ray sources
located in extra-nuclear regions of nearby galaxies with observed
X-ray luminosities exceeding the Eddington limit of
$\gtrsim10^{39}$~\lum\ for $\sim10M_{\sun}$ black holes. Recently,
evidence has been growing, primarily from \xmm\ high-quality X-ray
spectral and timing data
(\citealt{stobbart2006,gladstone2009,sutton2013}; see 
\citealt{feng2011,roberts2016} for recent reviews), that most ULXs are
stellar-remnant black holes \citep[with masses possibly reaching
  $\sim100~{\rm M}_{\sun}$][]{zampieri2009,belczynski2010} accreting
in super-Eddington, or `ultraluminous', regime
\citep{ss1973,poutanen2007}.

The salient ULX feature revealed by \xmm\ observations and witnessing
in favor of super-Eddington accretion is a turndown of the X-ray
spectrum above $\sim 5$--10~keV. However, as \xmm\ operates at
energies below $\sim 10$~keV, hard X-ray observations of ULXs were
highly anticipated to obtain better constraints on their spectra. The
first dedicated imaging hard X-ray observations of ULXs, namely M82
X-1 and Ho IX X-1, were performed in 2009--2013 with the IBIS
coded-mask instrument aboard {\it INTEGRAL} \citep{sazonov2014},
which, in combination with lower energy \xmm\ data, clearly revealed
a rollover above $\sim 10$~keV in the spectra of both sources. The
advent of the \textit{Nuclear Spectroscopic Telescope Array (NuSTAR)}
\citep{nustar} has opened a new era of hard X-ray observations of
ULXs. Its excellent angular resolution and broad (3--78~keV) energy
response made it possible for the first time to obtain a large set of
broad-band ULX spectra
\citep{W2013a,2014ApJ...793...21W,2015ApJ...799..122W,2015ApJ...806...65W,mukherjee2015,rana2015}.  

The source \x1\ (hereafter also referred to as \sn) in the nearby
Seyfert~2 galaxy NGC~5643 is one of the most luminous known ULXs, with
the X-ray luminosity of a few $10^{40}$~\lum. The source was
recently observed by \cite{annuar2015} in hard X-rays with \nu\ as
part of a campaign mainly devoted to local Compton-thick active
galactic nuclei. The hard X-ray (above 10~keV) emission from
\x1\ was measured for the first time and showed evidence for a
high energy spectral cutoff, thus confirming the ULX nature of the
source. 

The active nucleus of NGC~5643 (hereafter also referred to as the AGN) 
and \sn\ are separated by 52'', which is sufficent to spatially resolve
these X-ray sources with \nu, thanks to its 18'' FWHM angular
resolution. However, a detailed spectral analysis depends on the
amount of collected X-ray photons in a given area, which is commonly
expressed in terms of a half-power diameter (HPD), enclosing half of
the focused X-rays. Due to the wide wings of the \nu\ PSF, the
corresponding HPD reaches $\sim60''$ \citep{madsen2015}, which causes
partial confusion of the AGN and ULX. Therefore, \cite{annuar2015}
restricted the spectrum extraction areas around both sources to
circles of $20''$ radius. This, however, corresponds to only
$\sim$30\% PSF enclosed counts \citep{madsen2015} and thus implies a
significant loss in sensitivity. In this work, we utilize the full
collecting power of the \nu\ PSF to extract spatially resolved spectra
of the NGC~5643 nucleus and \sn, by applying 2D image fits to
\nu\ images in a number of narrow energy bands. This, together with
long \xmm\ observations of NGC~5643 taken in 2009 and 2014, enables us
to obtain high-quality broad-band spectra of \sn.

The paper is organized as follows. In Sect.~\ref{sec:data} we describe
the analysis of \nu\ (Sect.~\ref{sec:nustar}) and
\xmm\ data (Sect.~\ref{sec:xmm}). Variability and spectral analyses
are presented, respectively, in Sect.~\ref{sec:time} and
\ref{sec:spec}. The results are discussed and summarized in
Sect.~\ref{sec:discussion}. Following \cite{annuar2015}, we assume a
metric distance to NGC~5643 of $D=13.9$~Mpc \citep{sanders2003} based
on the cosmic attractor flow model described in \cite{mould2000}. We
assume that \sn\ is located in the galaxy NGC~5643.

\section{Observations and Data Analysis}
\label{sec:data}

\begin{table}
\noindent
\centering
\caption{List of X-ray observations of NGC~5643 used in this
  work}\label{tab:log} 
\centering
\vspace{1mm}
  \begin{tabular}{|c|c|r|r|c|c|c|}
\hline\hline
Mission & Date & ObsID & Exp. & { Frac.}$^{*}$ & Flux$^{**}$ \\
& & & (ks) & { (\%)} &  \\
\hline
{\it XMM} & 2009-07-25 & 0601420101 &  54.52 & { 87} & $2.32\pm0.10$ \\
{\it NuSTAR} & 2014-05-24 & 60061362002 & 22.46 & { 100} & $2.36\pm0.44$ \\
{\it NuSTAR} & 2014-06-30 & 60061362004 & 19.71 & { 100} & $2.45\pm0.50$\\
{\it XMM} & 2014-08-27 & 0744050101 & 116.9 & { 91} & $4.93\pm0.15$\\
\hline
\end{tabular}
$^*${ Good time fraction;} $^{**}$NGC~5643~X-1 (3--8~keV) unabsorbed flux in units of
  $10^{-13}$~\ergscm\ (see Sect.~\ref{sec:time} for details). 
\vspace{3mm}
\end{table}



\x1\ was first clearly detected in archival {\it ROSAT}/HRI
observations in 1997 by \cite{G04}. These authors analysed
\xmm\ observations of the Seyfert 2 galaxy NGC~5643 taken in 2003 and
found an X-ray source located $\simeq0.8'$ north-east of the nucleus,
which was 50 per cent brighter than the AGN. Under the assumption that
the source belongs to the NGC~5643 galaxy, its luminosity was
estimated at $\sim4\times10^{40}$~\lum, classifying \sn\ as an
ULX. New, longer \xmm\ observations were performed in 2009
\citep{matt2013} and 2014 \citep{pintore2016}. The first hard X-ray
($>10$~keV) imaging observations of NGC~5643 were carried out with
\nu\ in 2014 by \cite{annuar2015}, who also analysed simultaneous {\it
  Swift}/XRT and archival \xmm\ and {\it Chandra} observations
\citep{bianchi2006}. 

We use the \nu\ data to constrain the high-energy cutoff in the
spectrum of \x1\ and the long \xmm\ observations of 2009 and 2014 to
extend our spectral coverage to lower energies. The dates, exposures
and \textit{ObsID}s of these observations are listed in
Table.~\ref{tab:log}. 
 
Below we describe the procedures that we used to spatially separate the 
spectra of the NGC~5643 nucleus and \sn\ in the \nu\ data. We also
outline the standard data analysis for \xmm\ observations.

\subsection{\nu}
\label{sec:nustar}

NGC~5643 was observed with \nu\ in 2014 in two sessions separated by
37 days (Table~\ref{tab:log}). \nu\ carries two co-aligned twin X-ray
telescopes, with angular resolution of 18'' (FWHM) and HPD of
$\sim60''$, and operates in a wide energy range from 3 to 78~keV. The
detector modules of each telescope -- the focal plane modules A and B
(FPMA and FPMB) -- provide spectral resolution of $400$~eV (FWHM) at
$10$~keV. 

\begin{figure*}
\includegraphics[width=\textwidth]{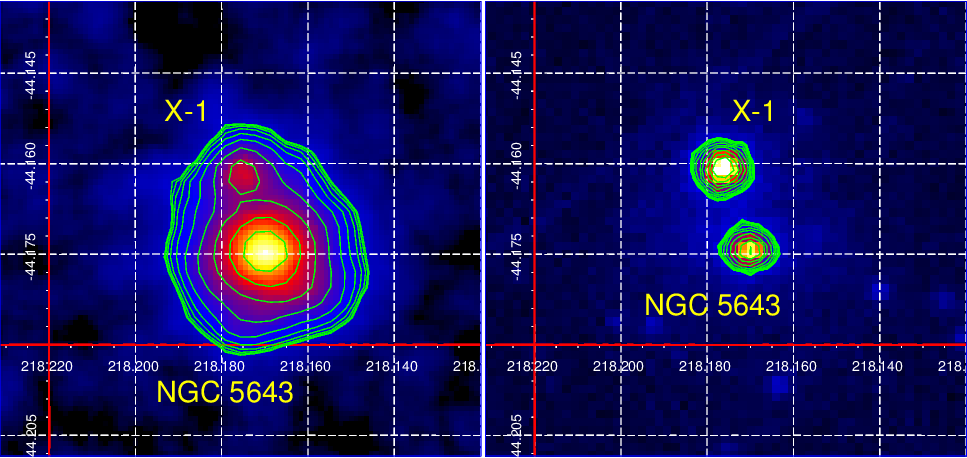}
\caption{{\it Left:} Exposure-corrected \nu\ mosaic image
  of NGC~5643 in the 3--78~keV band (May--June 2014). The image has been
  adaptively smoothed with the task {\it dmimgadapt} from {\sc
    ciao-4.7} using a tophat kernel. {\it Right:} Combined PN/MOS1/MOS2
  \xmm\ image of the same sky region in the 0.5--8~keV band (August
  2014). The green contours denote surface brightness in 10
  logarithmic steps from $0.1F_{\rm max}$ to $F_{\rm max}$, where
  $F_{\rm max}$ is the peak flux of the NGC~5643 nucleus in each
  image.}\label{fig:map}  
\end{figure*}

We processed the data from both modules using the NuSTAR Data Analysis
Software ({\sc nustardas}) v.1.5.1 and {\sc heasoft} v6.17. The data 
were filtered for periods of high instrumental background when the 
spacecraft passed the South Atlantic Anomaly and for known bad/noisy
detector pixels.

After running the pipeline, we ended up with event lists for the two
\nu\ observations with a total exposure of 42~ks
(Table~\ref{tab:log}). We performed astrometric correction of the
celestial coordinates of each incoming photon using NGC~5643
cataloged position and the centroid position of its AGN core in the
\nu\ images.


\begin{table}
\noindent
\centering
\caption{Best-fit parameters for the \x1\ spectrum measured with
  \nu, modelled by a power law modified by absorption of
  $10^{21}$~cm$^{-2}$ ({\it tbabs*powerlaw})}\label{tab:spe:nustar}
\centering
\vspace{1mm}
  \begin{tabular}{c|c|r|r}
\hline\hline
Parameter$^*$ & \multicolumn{3}{|c|}{\nu\ observations}\\
    &   60061362002 & 60061362004 & combined \\
\hline
$\Gamma$ & $3.06\pm0.50$ & $2.48_{-0.42}^{+0.51}$ & $2.82\pm0.35$ \\
$F_{3-24}$ &  $4.03\pm0.75$ & $4.34\pm0.95$ & $4.17\pm0.59$ \\
$\chi^{2}_{\rm r}$/d.o.f. & 1.15/13 & 0.94/13 & 1.23/13\\
\hline
\end{tabular}\\
$^*$The parameter $F_{3-24}$ represents the { 3--24~keV} unabsorbed flux
in units of $10^{-13}$~\ergscm.
\vspace{3mm}
\end{table}

\cite{annuar2015} pointed out that \x1\ did not show significant
variability between the two \nu\ observations. { To independently
  verify this assertion,} we extracted source spectra using 2D image
fitting procedures, described below, in order to estimate the
variability of \x1\ between the two \nu\ measurements in 2014. The
spectra were approximated by a power-law model modified by Galactic
and intrinsic absorption. As the energy response of \nu\ is not very
sensitive for measuring low absorption columns, we fixed the total
absorption column at $N_{\rm H}=10^{21}$~cm$^{-2}$, which is roughly
consistent with the results of the subsequent analysis involving \xmm\
low-energy data. The best-fit model parameters and derived {
  3--24~keV} fluxes are given in Table~\ref{tab:spe:nustar}. We see
that { within the uncertainties} the power-law slopes and
inferred fluxes are not significantly different between the two
observations { and consistent with those measured by
  \cite{annuar2015}. This justifies combining the two \nu\
  observations in 2014 for the following analysis.}  We conclude that
the available data are consistent with \x1\ being at the same
luminosity level during both \nu\ observations. We further discuss the
question of \x1\ variability using \xmm\ data in Sect.~\ref{sec:time}
below.

We finally combined the data from both \nu\ observations and both
\nu\ modules into sky mosaics in 15 energy bands logarithmically 
covering the \nu\ energy band of 3--78~keV. The subsequent analysis of
\nu\ data is based on these mosaic images and the broad-band spectra
derived from them. 

\begin{figure}
\includegraphics[width=0.98\columnwidth]{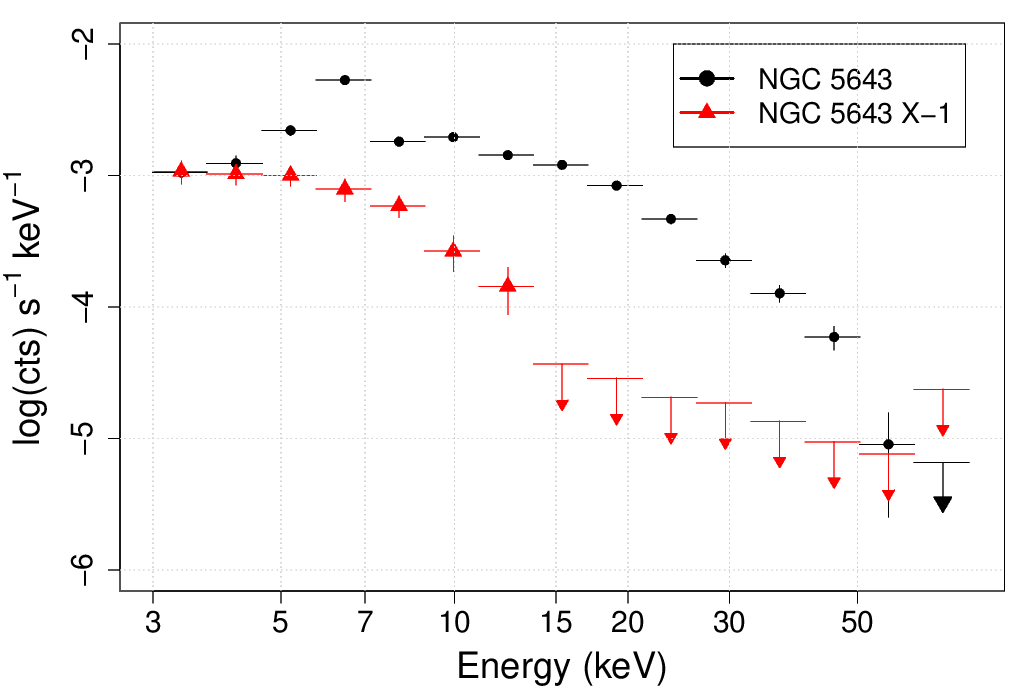}
\caption{Spatially decoupled \nu\ spectra of the NGC~5643 active
  nucleus { (black points)} and \x1\ { (red triangles)}. The
  upper limits are 1$\sigma$ errors. }\label{fig:spe}
\end{figure}


Fig.~\ref{fig:map} shows the combined image of NGC~5643 in the
3--78~keV energy band. Partial confusion of the AGN and \sn\ is
evident. To decouple emission from these two components, we fitted a
spatial model convolved with the \nu\ PSF using the {\sc sherpa}
package \citep{sherpa}, which is part of the {\sc ciao-4.7} software
\citep{ciao}. The spatial model of the sources is represented by two
2D Gaussians positioned at the NGC~5643 nucleus and \sn\ cataloged
coordinates. Following a similar analysis in \cite{arches}, the width
of the Gaussians were fixed at $4''$ FWHM, chosen to emulate the PSF
smearing effect. The amplitudes of the 2D Gaussians were free
parameters. { The \nu\ background is characterized by a spatial
  gradient across the field of view, unique to each mirror module
  \citep{wik2014}. However, as a result of co-adding of the FPMA and 
  FPMB data and data from the two observations with different spacecraft 
  position angles ($\Delta\alpha=40^{\circ}$), the background spatial 
  pattern has been largely averaged out and can be considered flat. 
  We thus estimated the constant background term at $R>250''$ from the AGN. 
  After running the fitting procedure in each of the 15 energy bands, we
  estimated the flux of both spatial components by sampling model
  parameter values over a Gaussian distribution with the {\it
    normal\_sample} command in {\sc sherpa}. The count rates and the
  corresponding errors, estimated from the characteristic values of
  the distribution ({\it plot\_cdf}), were combined} into PHA files
(denoted below with a '2D' superscript). { Note that the described
  2D-fitting gives the model flux (counts) for $\sim100\%$ \nu\ PSF
  enclosed energy, since we use a large fitting area ($R<250''$) around
  the NGC~5643 nucleus.}   
  To generate the corresponding RMF$^{\rm 2D}$ and
ARF$^{\rm 2D}$ response files, we ran the standard analysis with the
{\it nuproducts} tool for the AGN within the $R=30''$ circle,
producing a spectrum and response files for a point-like source. The
RMF matrix was used as is for RMF$^{\rm 2D}$. The effective area for
the spatially resolved spectra, ARF$^{\rm 2D}$, was estimated by
comparing the AGN 2D count rate (phot~s$^{-1}$, stored in PHA$^{\rm
  2D}$) and the corresponding count rate per cm$^2$ calculated in {\sc
  xspec} with the {\it flux} command (phot~s$^{-1}$~cm$^{-2}$) {
  operated on the model described below. Since the spectrum of the
  NGC~5643 nucleus reveals typical features of a Compton-thick AGN: a
  strong 6.4~keV iron line and a Compton reflection continuum above
  10~keV, we applied the {\it pexrav} model \citep{pexrav} commonly used
  for reflection-dominated spectra. The model also includes a primary
  power-law component with an exponential cutoff at 200~keV and  
  a slope $\Gamma$ tied to the corresponding {\it pexrav} model 
  parameter, and iron K$\alpha$ and K$\beta$ emission lines. The spectral
  modelling of the AGN in NGC~5643 is discussed in detail by \cite{annuar2015}
  and Krivonos et al. (2016, in preparation). Note that the ARF$^{\rm 2D}$
  created in this way is model dependent. We checked that different
  reflection-dominated models well fitting the spectrum of the NGC~5643 
  nucleus produce negligible deviations in ARF$^{\rm 2D}$.} 
  
  The 2D image analysis finally resulted in the spectra of the decoupled AGN 
  and \sn\ shown in Fig.~\ref{fig:spe}.

\subsection{\xmm}
\label{sec:xmm}

\xmm\ observations of NGC~5643 in 2009 and 2014 were performed with
the EPIC CCD cameras (PN, MOS1 and MOS2), operated in full frame mode
with the medium filter. We performed data processing with the {\sc
  sas} v1.2 data analysis software, applying time filtering for the
particle background.  

{ The data taken in 2009 were affected by
  soft proton flares known to be the main source of the high particle
  background for \xmm\ observations. \cite{pintore2016} filtered
  out 78\% of the data in 2009 due to background flare
  contamination. In our analysis, we considered \x1\ a moderately bright 
  X-ray source and assumed that it was not strongly affected by the flares. 
  The point-like shape of the source, in contrast to the extended faint
  sources, which are more affected by high particle background
  \citep{pradas2005}, provides support to this assumption. We 
  applied a moderate time filter $RATE<6.0$ (MOS1), $RATE<6.5$ (MOS2) and
  $RATE<50.0$ (PN), thus filtering out $\sim13\%$ of the data in 2009. We
  verified that stronger flare filtering conditions did not  
  significantly affected the resulting spectra of \x1. Time filters 
  $RATE<2.5, 3.0$ and $13.0$ were applied, respectively, to MOS1, MOS2 and 
  PN data taken in 2014, which contain only one clearly discernible proton 
  flare, filtering out only $\sim9\%$ of the total exposure.}  
  
  Spectra were extracted using a $R=25''$ aperture centred on the AGN 
  and \sn\ for all the cameras. Background spectra were extracted within 
  $R=100''$ circles located away from the sources, within the same chip. We used 
  patterns 0 to 4 and 0 to 12 for PN and MOS, respectively. { The right panel in
  Fig.~\ref{fig:map} demonstrates a PN, MOS1 and MOS2 combined
  0.5--8~keV mosaic image constructed from the 2014 data set. \x1\ appeared 
  on a CCD gap of the PN instrument during the observations in 2009. Therefore,
  to have a uniform data set and avoid dealing with multiple 
  cross-normalization coefficients, we excluded the 
  PN data from the spectral analysis and assumed there were no calibration 
  differences between the MOS1 and MOS2 cameras.}

\section{Time variability}
\label{sec:time}

To check if \x1\ varied over the considered period, { we fitted
  simultaneously the two \xmm\ (2009 and 2014) and two \nu\ (May 24 and
  June 30, 2014)} data sets with an { absorbed ({\it tbabs})}
cutoff power-law model, allowing free cross-normalizations between the
spectra. { This resulted in relatively poor fit statistics, 
  $\chi^{2}_{\rm r}$/d.o.f.= 1.13/695, indicating that the spectral model
  has not fully captured the data ($\chi^{2}_{\rm r}>1$) and suggesting 
  that spectral changes of \sn\ between the observations are possible. 
  The resulting best-fit model parameters $N_{\rm
    H}=(0.95\pm0.20)\times10^{21}$~cm$^{-2}$, $\Gamma=1.04\pm0.12$ and
  $E_{\rm cut}=5.7_{-1.0}^{+1.5}$ are close to those obtained for
  the high-quality data set taken in 2014, which likely dominates the
  fit.} The corresponding 3--8~keV flux estimates are listed in
Table~\ref{tab:log}. One can see that \sn\ brightened in the latest
(August 2014) \xmm\ observation by a factor of $\sim2$ compared to the
previous \xmm\ and \nu\ observations. This level of variability is
typical for this source, as indicated by its long-term light curve
presented in \cite{annuar2015} and \cite{pintore2016}.



\section{Spectral modelling}
\label{sec:spec}

\begin{figure*}
\includegraphics[width=\columnwidth]{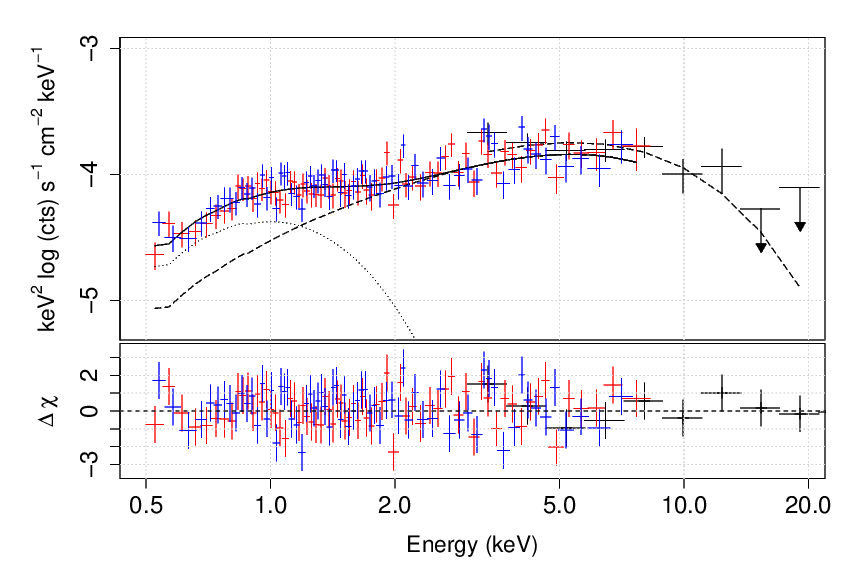}
\includegraphics[width=\columnwidth]{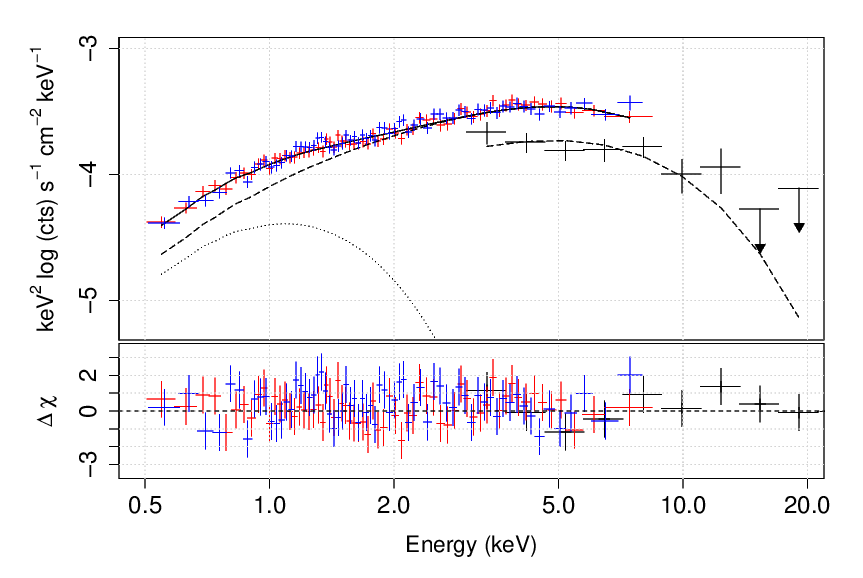}
\caption{{\it Upper panels:} X-ray spectra of \x1\  measured by
  \xmm\ (MOS1 and MOS2 spectra are shown in red and blue,
  respectively) in July 2009 (left) and August 2014 (right), and the 
  \nu\ spectrum measured in May--June 2014 (shown in black in both 
  panels). Different curves represent results of fitting by the
  \textit{tbabs*tbabs*const*(diskbb+compst)} model in {\sc xspec}. The
  full model (solid line) is shown for the \xmm\ data only;
  \textit{diskbb} and \textit{compst} are shown by dotted and dashed
  lines, respectively. {\it Lower panels:} the corresponding residuals
  of the data points and best-fit model in terms of
  $\Delta\chi$.}\label{fig:spe:diskbb}
\end{figure*}

{ In this section, we} describe our X-ray spectral analysis of the {
  \xmm\ (MOS1 and MOS2) and combined \nu\ data covering the 0.5--9~keV and
  3--78~keV energy bands, respectively. The MOS1 and MOS2 spectra measured
  in 2009 and 2014 are shown in Fig.~\ref{fig:spe:diskbb} together with the 
  \nu\ data points taken in 2014. Since \x1\ has not been detected above 20~keV, 
  we truncate the plot at this energy for a more detailed view. Note that 
  although we show $1\sigma$ upper limits for the two highest energy bins, 
  our spectral analysis described below uses the actual measurements.} Spectral 
  fitting and flux derivations were done using the {\sc xspec} package \citep{xspec},
photoionization cross sections from \cite{verner1996} and abundances
for the interstellar absorption from \cite{wilms2000}. We used $\chi^2$
statistics for fitting; all quoted errors are at the $90\%$ confidence
level.


As a first step in spectral modelling, we analysed the \nu\
data for \sn. We started with a simple power-law model modified
by line-of-sight absorption. The fitting yielded a photon index of
$\Gamma=2.82\pm0.35$ (Table~\ref{tab:spe:nustar}), with
$\chi^{2}=1.23$ for 13 degrees of freedom (d.o.f.). This slope measured
in the hard X-ray band is thus much steeper than the photon 
index $\Gamma\simeq1.6$ found in the soft X-ray band
\citep{G04,matt2013}, suggesting a high-energy downturn or a cutoff.

We then fitted the same absorbed power-law model to the \xmm\ data
taken in two epochs in 2009 and 2014. { As mentioned before, the
  available data do not rule out that \x1\ went through state changes
  over this period of time. Nevertheless, we combined the high-quality
  \xmm\ observations with non-simultaneous but unique \nu\ data to
  better constrain the high-energy part of the spectra where a cutoff
  is expected.}  An intrinsic absorption in the host galaxy was added
to the fixed Galactic absorption column density \citep[$N_{\rm
  H}^{Gal}=8\times10^{20}$~cm$^{-2}$,][]{kalberla2005} using two {\it
  tbabs} components. For the broad-band spectra, we also added a
cross-normalization constant between the \xmm\ and \nu\ data{,
  fixing the latter to unity.}  The measured best-fit model parameters
are shown in Table~\ref{tab:spe}. The fits are marginally acceptable.







\begin{table*}
\noindent
\centering
\caption{Best-fit spectral model parameters for \x1 measured with
  \xmm\ and \nu.}\label{tab:spe}
\centering
\vspace{1mm}
 \begin{tabular}{|c|c|r|r|r|c|c|}
\hline\hline
Parameter$^*$    & Unit & \multicolumn{2}{|c|}{\it XMM}  &
\multicolumn{2}{|c|}{\it XMM+NuSTAR}  \\
    & & 0601420101 &  0744050101 & 0601420101 &  0744050101 \\
    & & { Epoch 1} &  { Epoch 2} & { Epoch 1} &  { Epoch 2} \\
\hline
\multicolumn{5}{|c|}{Model = TBABS$\times$TBABS$\times$CONST$\times$POWERLAW}  \\
\hline
$N_{\rm H}$ & $10^{21}$~cm$^{-2}$ & $0.18_{-0.18}^{+0.40}$ & $1.36\pm0.22$ & $0.60\pm0.30$ & $1.16\pm0.22$ \\ 
$\Gamma$ & & $1.66_{-0.07}^{+0.09}$ & $1.60\pm0.04$ &
$1.78\pm0.10$ & $1.61\pm0.04$ \\
$N^{\rm pow}$& $\times10^{-5}$ & $7.80_{-0.55}^{+0.73} $ &
$17.64\pm0.74$ & $8.14_{-1.96}^{+2.46}$ &  $5.33\pm0.10$  \\
$C_{NuSTAR}^{XMM}$& & -- & -- & $0.98_{-0.17}^{+0.21}$ & $2.90_{-0.38}^{+0.50}$   \\
$\chi^{2}_{\rm r}$/d.o.f.& & 1.11/135 & 1.02/506 & 1.40/149  & 1.12/520  \\

\hline
\multicolumn{6}{|c|}{Model = TBABS$\times$TBABS$\times$CONST$\times$CUTOFFPL} \\
\hline
$N_{\rm H}$ & $10^{21}$~cm$^{-2}$ & $<8.22$ & $0.17_{-0.17}^{+0.36}$ & $<8.22$ & $<8.22$\\ 
$\Gamma$ & & $1.62\pm0.10$ & $0.95\pm0.14$ & $ 1.38\pm0.10$ & $0.88_{-0.08}^{+0.15}$ \\
$E_{\rm cut}$ & keV & $<500$ & $5.04_{-0.95}^{+1.78}$ &
$7.83_{-2.04}^{+3.42}$ & $4.50_{-0.67}^{+1.07}$\\
$N^{\rm cut}$& $\times10^{-5}$ & $7.81_{-0.30}^{+0.64}$ &
$16.76_{-0.36}^{+0.73}$ & $11.34_{-2.25}^{+2.63}$ & $8.62\pm1.30$  \\
$C_{NuSTAR}^{XMM}$& & -- & -- & $0.75_{-0.12}^{+0.15}$ &  $1.96_{-0.24}^{+0.30}$ \\
$\chi^{2}_{\rm r}$/d.o.f.& & 1.12/134 & 0.94/505 & 1.20/148 &  0.93/519 \\


\hline
\multicolumn{6}{|c|}{Model = TBABS$\times$TBABS$\times$CONST$\times$(DISKBB+CUTOFFPL)} \\
\hline
$N_{\rm H}$ & $10^{21}$~cm$^{-2}$ & $1.50_{-1.33}^{+2.01}$ & $1.02\pm0.45$ & $<2.77$ & $0.70_{-0.65}^{+0.79}$\\ 
$T_{in}$ & keV & $0.25_{-0.10}^{+0.15}$ & $0.33_{-0.08}^{+0.13}$ & $0.26_{-0.08}^{+0.11}$ & $0.31_{-0.07}^{+0.10}$ \\ 
$\Gamma$ & & $0.92_{-1.75}^{+0.66}$ & $0.12_{-0.85}^{+0.53}$ & $0.51_{-0.85}^{+0.56}$ & $0.36_{-0.57}^{+0.38}$ \\
$E_{\rm cut}$ & keV & $6.36_{-4.60}^{+6.46}$ & $2.57_{-0.77}^{+1.16}$ & $3.57_{-1.34}^{+2.54}$ & $2.91_{-0.76}^{+1.04}$\\
$N^{\rm bb}$& & $3.77_{-3.29}^{+42.30}$ & $1.03_{-0.68}^{+2.85}$ & $2.74_{-2.20}^{+17.02}$ &  $0.62_{-0.43}^{+1.71}$ \\
$N^{\rm cut}$& $\times10^{-5}$ & $5.83_{-3.08}^{+2.41}$ & $12.62_{-4.77}^{+3.13}$ & $6.00\pm0.30$ &  $7.17\pm2.00$ \\

$C_{NuSTAR}^{XMM}$& & -- & -- & $0.80_{-0.12}^{+0.15}$ &  $1.86_{-0.22}^{+0.28}$ \\

$L_{0.2-12}^{\rm bb}$& $10^{40}$\lum & $0.51_{-0.28}^{+0.96}$ &
 $0.53\pm0.22$ & -- & -- \\
$L_{0.2-12}^{\rm cut}$& $10^{40}$\lum &  $1.19\pm0.40$ &
 2.34$\pm0.35$ & -- & -- \\

$\chi^{2}_{\rm r}$/d.o.f.& & 1.04/132 & 0.92/503 & 0.98/146  &  0.90/517  \\

\hline
\multicolumn{6}{|c|}{Model = TBABS$\times$TBABS$\times$CONST$\times$(DISKBB+COMPST)} \\
\hline
$N_{\rm H}$ & $10^{21}$~cm$^{-2}$ & $1.33_{-1.18}^{+2.27}$ &
$0.93\pm0.25$ & $1.51_{-1.10}^{+2.00}$ & $1.22_{-0.54}^{+0.78}$ \\ 
$T_{in}$ & keV & $0.22_{-0.07}^{+0.15}$ & $0.31\pm0.18$ & $0.25\pm0.10$ & $0.28_{-0.17}^{+0.45}$ \\ 
$kT$ & keV & $2.01_{-2.01}^{+0.62}$ & $1.53\pm0.19$ & $1.85_{-0.32}^{+0.42}$ & $1.60_{-0.17}^{+0.18}$\\ 
$\tau$ & & $17.82_{-12.05}^{+15.34}$ & $23.49_{-3.09}^{+9.65}$ & $20.07_{-3.79}^{+7.73}$ & $22.13_{-2.40}^{+4.85}$ \\ 

$N^{\rm bb}$ &  & $4.80_{-4.39}^{+76.86}$ & $0.56_{-0.56}^{+0.60}$ &
$3.41_{-4.42}^{+58.13}$ & $0.32_{-0.30}^{+6.09}$ \\ 
$N^{\rm comp}$ & $\times10^{-5}$  & $6.50_{-3.70}^{+2.03}$ &
$14.28_{-1.85}^{+1.90}$ & $7.51_{-2.99}^{+2.88}$ & $8.02_{-2.33}^{+1.47}$ \\ 

$C_{NuSTAR}^{XMM}$& & -- & -- & $0.82_{-0.13}^{+0.16}$ &  $1.87_{-0.22}^{+0.29}$\\

$L_{0.2-12}^{\rm bb}$& $10^{40}$\lum & $0.48_{-0.29}^{+1.22}$ &
$<0.21$ & -- & -- \\
$L_{0.2-12}^{\rm comp}$& $10^{40}$\lum &  $1.31\pm0.40$ &
2.80$_{-0.23}^{+0.35}$ & -- & -- \\

$\chi^{2}_{\rm r}$/d.o.f.& & 1.05/132  & 0.92/504  & 1.02/146  & 0.92/517  \\

\hline
\multicolumn{6}{|c|}{Model = TBABS$\times$TBABS$\times$CONST$\times$(DISKBB+DISKPBB)} \\
\hline
$N_{\rm H}$ & $10^{21}$~cm$^{-2}$ & $1.31_{-1.30}^{+1.80}$ & $0.60_{-0.60}^{+0.80}$ & $1.27_{-0.12}^{+1.70}$ & $0.69_{-0.64}^{+0.84}$\\ 
$T_{in}^{\rm bb}$ & keV & $0.24_{-0.07}^{+0.11}$ & $0.33_{-0.09}^{+0.11}$ & $0.25_{-0.07}^{+0.11}$ & $0.31_{-0.09}^{+0.11}$ \\ 
$T_{in}^{\rm pbb}$ & keV & $3.02_{-3.02}^{+1.24}$ & $1.95_{-0.25}^{+0.40}$ & $2.53_{-0.59}^{+0.88}$ & $2.10_{-0.30}^{+0.38}$ \\ 
$p$ & & $0.62_{-0.62}^{+0.08}$ & $0.78_{-0.78}^{+0.13}$ & $0.66_{-0.08}^{+0.34}$  & $0.72_{-0.08}^{+0.27}$ \\
$N^{\rm bb}$ &  & $3.31_{-2.95}^{+40.57}$ & $0.87_{-0.60}^{+2.89}$  &
$1.55\pm0.62$ & $0.62\pm0.35$ \\ 
$N^{\rm pbb}$ & $\times10^{-3}$  & $0.15_{-0.12}^{+2.80}$ &
$3.63_{-2.47}^{+5.77}$ & $1.01_{-0.74}^{+3.05}$ & $1.28_{-0.72}^{+1.48}$ \\ 
$C_{NuSTAR}^{XMM}$& & -- & -- & $0.80_{-0.12}^{+0.16}$ &  $1.84_{-0.22}^{+0.28}$ \\
$L_{0.2-12}^{\rm bb}$& $10^{40}$\lum & $0.49_{-0.24}^{+1.12}$ &
$0.47_{-0.24}^{+0.21}$ & -- & -- \\
$L_{0.2-12}^{\rm pbb}$& $10^{40}$\lum &  $1.20_{-0.32}^{+0.35}$ &
$2.40_{-0.23}^{+0.35}$ & -- & -- \\
$\chi^{2}_{\rm r}$/d.o.f.& & 1.04/132 & 0.92/503 & 0.98/146   &  0.90/517  \\

\hline
\end{tabular}\\
$^*$The model parameters for the power-law, power-law with an
exponential cutoff ({\it cutoffpl}), multi-temperature blackbody
accretion disk ({\it diskbb}), a Comptonization spectrum ({\it compST}), and
multi-temperature disk blackbody with power-law dependence of
temperature on radius ({\it diskpbb}), are shown with $N^{\rm pow}$,
$N^{\rm cut}$, $N^{\rm bb}$,  $N^{\rm comp}$, and $N^{\rm pbb}$, respectively.
\vspace{3mm}
\end{table*}

 As mentioned above, the hard X-ray data indicate the presence of a
 high-energy cutoff (see also \citealt{annuar2015}). We thus next
 fitted an absorbed cutoff power-law model ({\it cutoffpl} in {\sc
   xspec}). This led to a significant improvement in the fit
 statistics and allowed us to constrain the cutoff energy for the
 combined \xmm\ and \nu\ spectra at $E_{\rm
   cut}=7.83_{-2.04}^{+3.42}$~keV and $4.50_{-0.67}^{+1.07}$~keV for
 the first and second epochs, respectively.  { The former value
   agrees with and significantly improves upon the $E_{\rm
     cut}=16.3_{-9.6}^{+35.9}$~keV constrain obtained by
   \cite{annuar2015} using the same 2009 \xmm\ and 2014 \nu\ data
   sets.  Regarding the second epoch, $E_{\rm cut}$ is consistent with
   \cite{pintore2016} result $E_{\rm cut}=5.4_{-0.8}^{+1.0}$ based on
   \xmm\ only data taken in 2014, which is also seen in
   Table~\ref{tab:spe}. Here, the \nu\ data further narrow the allowed
   range for the cutoff energy. The $\Gamma$ and $E_{\rm cut}$
   parameters measured for the two epochs using a combination of \xmm\
   and \nu\ data are marginaly consistent with each other and suggest a change
   in the spectral state of \sn.}


ULX spectra are often described with { a sum of two spectral components}: 
hard X-ray emission, presumably emergent from the
inner regions of a supercritical accretion disc, and soft X-ray emission, 
often attributed to a thick outflowing wind. We thus next fitted a
two-component, {\it diskbb} plus power-law, model (with absorption),
but only to the \xmm\ data. { The best-fit parameters are listed in
Table~\ref{tab:spe:xmm}.} This was primarily done to determine which
type \x1\ might belong to in the ULX classification scheme introduced by
\cite{sutton2013} (see Sect.~\ref{sec:discussion} below). The inner disc 
temperature of { the {\it diskbb} component} in the 2014 data was not 
constrained by the fit, and so we fixed it at 0.3~keV to be consistent with 
the following spectral analysis. { The inferred power-law photon index $\Gamma<2$
(both in 2009 and in 2014) and low multicolour-disc temperature $kT_{\rm in}<0.5$~keV 
{ (reliably measured in the 2009 \xmm\ observation)} put \x1\ into the 
class of {\it hard ultraluminous} ULX states.}

\begin{table}
\noindent
\centering
\caption{ Best-fit spectral model parameters for \x1 measured with
  \xmm\ modelled in {\sc xspec} notation as {\it
    tbabs*tbabs*(diskbb+powerlaw)}. The unabsorbed $0.2-12$~keV flux of {\it
    diskbb} and {\it powerlaw} components are shown with
  $F_{0.2-12}^{\rm bb}$ and $F_{0.2-12}^{\rm pow}$, respectively. }\label{tab:spe:xmm}
\centering
\vspace{1mm}
 \begin{tabular}{|c|c|r|r|r|c|c|}
\hline\hline
Parameter    & Unit & \multicolumn{2}{|c|}{\it XMM}    \\
    & & 0601420101 &  0744050101  \\
    & & {Epoch 1} &  {Epoch 2} \\

\hline
$N_{\rm H}$ & $10^{21}$~cm$^{-2}$ & $2.00_{-1.40}^{+1.18}$ & $1.36\pm0.22$ \\ 
$T_{in}$ & keV & $0.21_{-0.05}^{+0.10}$ & 0.3 (fixed)  \\ 
$\Gamma$ & & $1.55\pm0.17$ & $1.59\pm0.04$  \\
$F_{0.2-12}^{\rm bb}$& $10^{-13}$\ergscm & $2.32_{-1.50}^{+5.20}$  & $<1.00$     \\
$F_{0.2-12}^{\rm pow}$& $10^{-13}$\ergscm & $6.34\pm0.45$  & $15.49\pm0.28$    \\

$\chi^{2}_{\rm r}$/d.o.f.& &  1.04/133  & 1.02/505 \\

\hline
\end{tabular}\\
\vspace{3mm}
\end{table}

We then replaced the power law for the hard component by a high-energy
cutoff power-law model and applied the resulting {\it diskbb} plus
{\it cutoffpl} model to the \xmm\ and combined \xmm\ and \nu\
data. { This significantly improved the fit statistics for the
  first epoch, both for the \xmm\ data alone and for the combined 
  \xmm\ and \nu\ data, with respect to single power-law and {\it cutoffpl} 
  models (Table~\ref{tab:spe}), which suggests a two-component structure 
  of the ULX spectrum. Note that \cite{pintore2016} could fit the \sn\ 
  spectrum well with just one broad component, analyzing the same \xmm\ data 
  set taken in 2009. We determined the statistical significance of the soft 
  {\it diskbb} component using the {\sc xspec} script {\it simftest} with 
  $3\times10^4$ trials, and found that the null hypothesis, i.e. no 
  soft component, can be rejected at a $3.3\times10^{-3}$ significance level, 
  which corresponds to a $\sim3\sigma$ detection of the soft component, 
  assuming a normal distribution. We can speculate that the 
  non-detection of the soft component by \cite{pintore2016} might have been 
  mainly caused by their too strong data filtering (see Sect.~\ref{sec:xmm}).  
  The spectrum of the second epoch is dominated by the hard component, and
  the addition of a soft multicolour-disc component provides only a marginal 
  improvement (as also mentioned by \citealt{pintore2016}). The corresponding
  significance of the soft component estimated as described above is
  $3.2\sigma$.} 
  
  { We conclude that the {\it diskbb} plus {\it cutoffpl} phenomenological 
  model constrains well both the {\it diskbb} temperature and the shape of the 
  hard component, and implies a low $kT_{\rm in}<0.5$~keV, hard photon index 
  $\Gamma\sim0.5$ and relatively low $E_{\rm cut}\sim 2$--6~keV. Note that \nu\ 
  data have low sensitivity to the soft component due to the energy response 
  starting at 3~keV. Hence, we do not provide estimates of the luminosities of 
  the spectral components in the 0.2--12~keV band in Table~\ref{tab:spe} for 
  joint \xmm\ and \nu\ data, as we do for \xmm\ data alone.} 

Power-law spectra with a high-energy cutoff can be explained as
Comptonization of soft radiation by hot electrons. We thus next tried
to describe the hard component with the \textit{compst} model for
Comptonized radiation from a cloud of hot gas \citep{sunyaev1980}. The
resulting gas temperature is $\sim2$~keV and the Thompson optical
depth of the cloud is $\sim$20, similar to findings for other ULXs
(e.g. \citealt{sazonov2014,kobayashi2016}). In all other respects,
this model is almost indistinguishable from the previous, {\it diskbb}
plus {\it cutoffpl}, model. 

We finally fitted the \xmm\ and \nu\ spectra with the multicolour
blackbody accretion disc model \textit{diskpbb} with a variable
temperature disc profile \citep{diskpbb}{, in combination with a 
soft \textit{diskbb} component.} The fit statistics are
similar to those obtained for the preceeding two models. The inferred 
radial temperature profile parameter $p$ is smaller than expected for
a thin disc \citep[$p<3/4$,][]{ss1973}, which is consistent with a
radiation dominated accretion disc.

\section{Discussion and summary}
\label{sec:discussion}

The two long \xmm\ observations of \x1\ taken in 2009 and 2014 allow 
us to compare its spectral states in two epochs separated by five
years. {The \nu\ data taken 3 and 2 months} before the later
\xmm\ observation{, and analysed here paying special attention to 
the separation of the contributions of the ULX and the nearby active nucleus of 
NGC~5643,} provide a more stringent constraint on the high-energy cutoff 
and further information on the luminosity state of the source.

Our analysis of the \xmm\ and \nu\ data has demonstrated that the spectrum 
of \x1\ contains two components: i) a dominant hard one, which can be 
empirically described as a power law with photon index $\Gamma\sim0.5$ 
and a cutoff at $E_{\rm cut}\sim 2$--6~keV (fitting the \xmm\ data
by a simple power-law model yields $\Gamma\approx 1.6$), and ii) a 
soft one, characterized by temperature $kT_{\rm in}\sim 0.3$~keV 
if described in terms of multicolour disc emission. These spectral 
parameters suggest that \x1\ was in the {\it hard luminous} ULX state 
during these observations, according to the \cite{sutton2013} empirical 
classification scheme ($kT_{\rm in}<0.5$~keV, $\Gamma<2$). 

{ 
\cite{pintore2016} recently carried out a spectral and temporal 
analysis of all the available \xmm\ observations of \x1\ and suggested  
that its spectrum could be well modelled by a single broad, thermal-like 
component, such as an advection dominated disc or an optically thick 
Comptonizing corona. In their analysis, the {\it diskpbb} model fitted the 
data better than a combination of a soft {\it diskbb} and power-law 
components, suggesting that the source might belong to the 
{\it broadened disc} class rather than the {\it hard luminous} class in the 
empirical ULX scheme of \cite{sutton2013}. Nevertheless, \cite{pintore2016} 
favoured association of \x1\ with the latter class given the high X-ray 
luminosity of the source, and in fact demonstrated that a sum of a hard 
Comptonization component and a soft {\it diskbb} component led to a somewhat 
better fit compared to a single broadened disc model. These conclusions were 
mostly based on the 2014 \xmm\ data set, for which we obtain very similar results, 
i.e. the soft {\it diskbb} component is only marginally detected in these data. 
However, according to our new analysis, the \xmm\ data clearly reveal the 
presence of an additional, soft $\sim0.3$~keV thermal-like component in the 2009
data set, which was only briefly discussed by \cite{pintore2016} since they 
had filtered out the bulk of the \xmm\ data for that observation. 
}

The 0.2--12~keV luminosity of the source nearly doubled in the second
\xmm\ observation compared to the first one, from $\sim1.5\times10^{40}$ 
to $\sim3\times10^{40}$~\lum. However, this change in luminosity 
apparently reflects \x1\ variability on a time-scale of weeks rather
than years, since both \nu\ observations took place {within three months} 
before the \xmm\ observation {in August 2014} and found the source at the
same (low) luminosity level as in the \xmm\ observation taken in
2009. Most if not all of this variability can be attributed to the
hard spectral component. The luminosity of the soft component is
consistent with being constant but the data permit it to have varied by
a factor of $\sim 2$ between the observations. We have not detected
any significant changes in the shape of either spectral component. 

Similar to other ULXs with high-quality data, the observed
two-component spectrum of \x1\ can be interpreted in terms of
supercritical accretion on to a stellar-mass black hole in a binary
system. In this scenario \citep{ss1973,poutanen2007,middleton2015}{, previously
discussed in application to \x1\ by \cite{pintore2016},} 
the hard component is emission from the hot, inner regions of a thick
accretion disc { observed nearly face-on}, collimated by reflections 
off the walls of its central funnel, while the soft component is photons 
diffusing out from a massive wind outflowing from the accretion disc, 
which is possibly somewhat collimated too \citep{king2009}. If so, the 
two-fold increase in the hard X-ray luminosity of \x1\ observed by \xmm\ in 
August 2014, as compared to the previous \xmm\ and \nu\ observations, may be 
primarily the result of an increased beaming (i.e. decreased opening angle of the
funnel) caused by a moderate increase in the accretion rate on to the
black hole.

\centerline{}

\section*{Acknowledgments}

This work has made use of data from the \nu\ mission, a project
led by the California Institute of Technology, managed by the Jet
Propulsion Laboratory and funded by the National Aeronautics and
Space Administration, and reobservations obtained with XMM-Newton, an
ESA science mission with instruments and contributions directly funded
by ESA Member States and NASA. The research has made use of the NuSTAR Data
Analysis Software ({\sl nustardas}) jointly developed by the ASI
Science Data Center (ASDC, Italy) and the California Institute of
Technology (USA). The work was supported by the Russian Science
Foundation (grant 14-12-01315).




\label{lastpage}

\end{document}